# Consequences of cell-to-cell P-glycoprotein transfer on acquired multidrug resistance in breast cancer: a cell population dynamics model

Jennifer Pasquier[†*], Pierre Magal[‡*], Céline Boulangé-Lecomte[†], Glenn Webb[¶] & Frank Le Foll[†§]

[†]*Laboratory of Ecotoxicology UPRES EA 3222, IFRMP 23, University of Le Havre, 76058 Le Havre cedex, France*
[‡]*UMR CNRS 5251 IMB & INRIA sud-ouest Anubis, Université de Bordeaux, 33076 Bordeaux, France*
[¶]*Department of Mathematics, Vanderbilt University, 1326 Stevenson Center, Nashville, TN37240.*
[*]*These authors have contributed equally to the studies presented in this manuscript*

## ABSTRACT

*Background*

Cancer is a proliferation disease affecting a genetically unstable cell population, in which molecular alterations can be somatically inherited by genetic, epigenetic or extragenetic transmission processes, leading to a cooperation of neoplastic cells within tumoral tissue. The efflux protein P-glycoprotein (P-gp) is overexpressed in many cancer cells and has known capacity to confer multidrug resistance to cytotoxic therapies. Recently, cell-to-cell P-gp transfers have been shown. Herein, we combine experimental evidence and a mathematical model to examine the consequences of an intercellular P-gp trafficking in the extragenetic transfer of multidrug resistance from resistant to sensitive cell subpopulations.

*Methodology and principal findings*

We report cell-to-cell transfers of functional P-gp in co-cultures of a P-gp overexpressing human breast cancer MCF-7 cell variant, selected for its resistance towards doxorubicin, with the parental sensitive cell line. We found that P-gp as well as efflux activity distribution are progressively reorganized over time in co-cultures analyzed by flow cytometry. A mathematical model based on a Boltzmann type integro-partial differential equation structured by a continuum variable corresponding to P-gp activity describes the cell populations in co-culture. The mathematical model elucidates the population elements in the experimental data, specifically, the initial proportions, the proliferative growth rates, and the transfer rates of P-gp in the sensitive and resistant subpopulations.

*Conclusions*

We confirmed cell-to-cell transfer of functional P-gp. The transfer process depends on the gradient of P-gp expression in the donor-recipient cell interactions, as they evolve over time. Extragenetically acquired drug resistance is an additional aptitude of neoplastic cells which has implications in the diagnostic value of P-gp expression and in the design of chemotherapy regimens

## 1. INTRODUCTION

*Current view on cancer*

In essence, cancer is a proliferation disease affecting a genetically unstable cell population. The somatic mutation theory, which is the prevailing paradigm, stipulates that cancer arises from a stepwise accumulation of changes in genes that progressively drives subclonal neoplastic cells to evolve independently from the others, escaping from proliferation control and competing for space and resources and, finally, to kill the host. The key feature is that the cancer-promoting changes are intrinsic molecular events that are *i)* inheritable by daughter cells and *ii)* selectable, in the sense that they confer selective advantage in the cell population environment [1, 2]. It is also widely recognized that genetic instability results in the creation of diverse daughter cells, capable of interaction with the cell microenvironment, conferring to the tumor the properties of a heterogeneous tissue [3].

Within such tumor tissue, molecular alterations are considered to be somatically inherited by genetic, epigenetic (traits that are not dependent on the primary sequence of DNA) or extragenetic (post-translational protein modifications that initiate and support positive feedback loops) transmission processes [1], leading to a cooperation of neoplastic cells [4].

*Role of P-glycoprotein in breast cancer*

Despite decades of research, breast cancer remains a major public health issue. Worldwide, breast cancer represents 23 % of overall female cancers and accounts for 40,000 deaths each year in the United States alone. Currently, the lifetime risk of developing a breast cancer for women born in the United States is 1 in 8. For metastatic forms of breast cancers, the 5-year relative survival is only 26 % [5] and complete remission has been estimated as low as 3.1 % after this period [6]. Response to first-line chemotherapies span from 30 to 65 % and are followed, after a period varying from 6 to 10 months, by disease progression [7]. In fact, resistance to chemotherapy is believed to cause treatment failure in 90 % of metastatic breast cancer patients [8].

Resistance to chemotherapy is related to the overall mechanisms that are involved in a decrease of drug efficacy upon tumors [9]. Among these factors, proteins lowering the intracellular concentration of chemotherapeutics and belonging to the ATP-binding

---

Cassette (ABC) transporters are well-known for their specific responsibility [10]. These membrane proteins are characterized by their ability to efflux a large panel of both chemically and functionally unrelated compounds comprising potent cytotoxics currently used in chemotherapeutic treatments. Consequently, tumors overexpressing this type of energy-dependent pump have been very early identified on the basis of their *Multi-Drug Resistance* (MDR) phenotype [11]. The P-glycoprotein (P-gp) was the first drug-efflux protein characterized [12]. With the growing number of molecularly cloned ABC transporters (48 different genes up to now have been identified in the human genome), a rational nomenclature have been proposed. Meta-analysis [13] or immunochemistry studies [14] have determined that approximately 40 % of all breast cancer tumors express *ABCB1/MDR1* coded P-gp. *ABCB1/MDR1* gene expression has a prognostic value for cell resistance to anticancer drugs [15, 16] and for treatment failure [17].

*Genetic versus extragenetic resistance transmission*

Until now, the development of multi-drug resistance in neoplastic cells was explained as the consequences of two main mechanisms. On one hand, cells natively expressing drug-efflux proteins conserve their phenotype throughout the process of malignant transformation. On the other hand, in non-expressing cells, chemotherapeutics have been shown first to induced P-gp expression and, second, to exert a selection of resistant cells during the course of chemotherapy [18-21]. Therefore, P-gp induction was considered to be only dependent on the cell type and/or the previous history of exposure to cytotoxics. Biomathematical models have been proposed to describe the kinetics of P-gp induction as a function of tissue cytotoxic concentration [22] with the objective of adapting the time course of drug administration to overcome as much as possible multidrug resistance.

Recently however, extragenetic transmissions of multidrug resistance have been reported. These unexpected events all involve direct transfers of P-gp to *ABCB1/MDR1* non-expessing recipient cells from various multidrug resistance donors, namely adherent cell lines *in vitro* and *in vivo* [23], stromal cells isolated from patients suffering of ovarian cancers [24] and lymphblastic leukemia cells in suspension [25]. These findings actually add a new modality of resistance appearance, and possibly spreading, in a population of tumor cells. Theoretically, cell-to-cell P-gp transfers in conditions of antineoplastic treatment could confer a significant advantage to sensitive cells, keeping them alive long enough to produce their own P-gp copies, under the induction process, and thus, to resist chemotherapy. However, the real dynamics of extragenetic acquired resistance in a growing population of cancer cells remains, up to now, unknown. Herein, we propose a biomathematical model, derived from a previous analysis [26], supported with experimental parameters and combining processes of *i)* cell proliferation and death, *ii)* P-gp induction and degradation and *iii)* cell-to-cell P-gp transfer, to help answer this question. The present model was supplied by new data obtained *in vitro* from MCF-7 human breast adenocarcinoma and capable of intercellular P-gp transfer [23]. Comprehensive mechanistic aspects of cell-to-cell P-gp transfers are presented in a companion paper [27]. The main objectives of the present work were to investigate the influence of both cancer cell line intrinsic factors (growth rate, initial proportions of sensitive and resistant cells, and P-gp transfer rate), as well as treatment management parameters (*i.e.* cycles of drug administration), on overall multidrug resistance, both genetically and extragenetically transmitted.

## 2. MATERIAL AND METHODS

*2.1. Cell lines.*

The cell lines used in the present study were the wild-type drug-sensitive human breast adenocarcinoma MCF-7, purchased at the American Type Culture Collection, and a multi-drug resistant MCF-7/DOXO variant, kindly provided by Pr. J.-P. Marie (Hôtel Dieu, Paris, France). The cells were grown in RPMI 1640 medium supplemented with 5% heat-inactivated fetal bovine serum, 2 mM L-glutamine, and 1% antibiotic/antimycotic solution. Doxorubicin (1 µM) was added to the culture medium for the maintenance of the multi-drug resistant phenotype of MCF-7/doxo cells. Cultured cells were incubated at 37°C under a water-saturated 95% air-5% $CO_2$ atmosphere. Exposure of MCF-7/DOXO to the P-gp inhibitors verapamil or cyclosporine A dose-dependently abolished both resistance to doxorubicin and P-gp activity in MCF-7/doxo [28].

For proliferation assays, cells were plated at $10^5$ per 60 X 15 mm tissue culture dish with or without doxorubicin. Cells were dissociated by treatment with trypsin/EDTA and then counted in a Malassez chamber. Cell proliferation was followed every 12 hours during six days.

*2.2. Analysis of P-gp expression and activity by flow cytometry.*

Cultured or co-cultured MCF-7 were trypsin-resuspended and washed with HBSS before analysis. For intercellular P-glycoprotein transfer studies, P-gp was labelled using a phycoerythrin (PE)-conjugated UIC2 mAb (Beckman Coulter France, Villepinte, France). The fluorescent light (FL) was quantified using a Cell Lab Quanta SC MPL flow cytometer (Beckman Coulter) equipped with a 22 mW 488 nm excitation laser. Voltage settings of photomultipliers were not modified throughout the experiments. Each analysis consisted in a record of 10 000 events, triggered on electronic volume (EV) as primary parameter, according to a particle diameter exceeding 8 µm. Red fluorescence was measured in FL2 channel (log scale) through a 575 nm band pass emission filter. More than 93.1 ± 0.4 % (mean ± S.E.M.) of gated events exhibited a FL2>1 for MCF-7/DOXO.

To study P-gp activity, resuspended cells were loaded with 0.25 µM calcein acetoxy-methylester (Invitrogen Life Technologies, Carlsbad, CA) in RPMI for 15 min at 37°C in the dark. Green FL was quantified via the FL1 channel (log scale) through a 525 nm band pas filter. Controls of MCF-7, MCF-7/DOXO and extemporaneous mixtures of 50:50 MCF-7: MCF-7/doxo were analysed before co-cultures, in all experiments.

### 2.3. Analysis of P-gp by flow cytometry transfer in tagged MCF-7.

75 %-confluent sensitive MCF-7 were loaded with 25 µM of the cell-permeant reactive fluorescent dye CellTracker Blue 4-chloromethyl-6,8-disulforo-7-hydroxucoumarin (Invitrogen Life Technologies, Carlsbad, CA) for 2 hour at 37°C in 25 cm² culture dishes. The cells are then washed twice with HBSS and covered with fresh complete culture medium. Either CellTracker Blue loaded sensitive MCF-7 (ctbMCF-7), or MCF-7/doxo, or a carefully homogenized 50:50 mixture of ctbMCF-7:MCF-7/doxo were plated at the desired density on 60mm culture dishes and analysed by flow cytometry after labelling using the PE-conjugated UIC2 mAb, as described before.

### 2.4. Flow cytometry data conditioning.

Accurate fitting of the mathematical model to flow cytometry all-events histograms required raw data noise reduction. Thus, list-mode data were segregated into binned logarithmic histograms, with a bin width of 0.05 in log scale of arbitrary fluorescence units. The distributions were normalized, by scaling to an integral area of 1.0, in order to obtain a density of probability.

### 2.5. Mathematical model of P-gp activity transfer.

In this section, we describe a mathematical model that has been developed and used to describe the consequences of a cell-to-cell P-gp protein transfer in terms of multidrug resistance activity. Herein, the resistance activity is viewed as the ability of a cell to efflux cytotoxics, which is considered to be correlated to the number of P-gp copies within the cell membrane. We decompose the construction of the model by considering separately the process of *i)* cell proliferation and death, *ii)* P-gp induction and degradation rate, both, at the cell and the population level, and *iii)* P-gp transfer between cells. We consider $p$ the P-gp efflux activity of a cell, functionally measured by the loss of calcein fluorescence for a resistant cell compared to a sensitive one, and we introduce $u(t,p)$ the density of cells at time $t$ having a P-gp activity $p$. In the sequel, for a given cell, the fluorescence $p$ is assumed to be a function of P-gp expression on the cell surface.

The terminology "density of cells with respect to $p$" means that, if $p_1 < p_2$ are two quantities of P-gp activity, then the number of cells with a P-gp activity in between $p_1 < p_2$ at time $t$ is

$$\int_{p_1}^{p_2} u(t,p)dp.$$

According to the fluorescence scale used in cytometry, $p$ varies between $p_{min}=1$ and $p_{max}=10^4$, and the total number of cells is given by

$$U(t) = \int_{p_{min}}^{p_{max}} u(t,p)dp.$$

In order to describe the time course of the cell distribution, we introduce the following model which has already been described in [26]

$$\frac{\partial u(t,p)}{\partial t} = \underbrace{\varepsilon^2 \frac{\partial^2 u(t,p)}{\partial p^2}}_{\text{stochastic drift}} - \underbrace{\frac{\partial}{\partial p}(hu(t,p))}_{\text{P-gp activity drift}} + \underbrace{\rho(p)u(t,p)}_{\text{proliferation}} + \underbrace{2\tau(T(u(t;\cdot))-u(t,p))}_{\text{P-gp transfer}}$$

$$\varepsilon^2 \frac{\partial u(t,p)}{\partial x} - h(p)u(t,p) = 0, \text{ for } p = p_{min}=1 \text{ and } p = p_{max}=10^4$$

$$u(0,p) = u_0(p) \in L^1_+(1,10^4)$$

In this model, the diffusion term describe a stochastic drift of P-gp activity (with no bias), while the convection term describes an oriented drift of the activity of P-gp, and the proliferation term combines the cell division and mortality of cells, which is assumed to depend on the activity $p$ of P-gp only during the treatment. Robin type boundary conditions are introduced to preserve the total number of individuals when only the drift process takes place (i.e. $\rho=0$ and $\tau=0$). The term describing transfer of P-gp has been presented in detail in [26]. The parameter $\tau$ describes the time between two consecutives transfers, and the transfer of operator $T$ is defined by

$$T(\varphi)(p) = \frac{\int_{-\infty}^{+\infty} \overline{\varphi}(p + f(\hat{p})\hat{p})\overline{\varphi}(p - (1-f(\hat{p}))\hat{p})d\hat{p}}{\int_1^{10^4} \varphi(\hat{p})d\hat{p}}, \phi \in L^1_+(1,10^4),$$

where $\overline{\varphi}$ extends $\varphi$ by 0 outside of the interval $(1,10^4)$. The main idea in the construction of the transfer operator $T$ is to derive the probability of a recipient partner to acquire a level $p$ of P-gp in a transfer event from all possible donor partners. To describe a transfer event we use the following rule: if a cell $C_1$ and a cell $C_2$ have, respectively, a quantity $p_1$ and $p_2$ of P-gp activity before transfer, then, after transfer, $C_1$ (respectively, $C_2$) will have an activity $p_1-f(|p_1-p_2|)(p_2-p_1)$ (respectively, $p_2-f(|p_1-p_2|)(p_1-p_2)$ ). So the fraction transferred *is $f(|p_1-p_2|)$*, which depends on the absolute value of the difference between $p_1$ and $p_2$ (*the distance between the P-gp activities of $C_1$ and $C_2$*). The model is used with constraints that correspond to a permissible interval ($\delta_{min}, \delta_{max}$) for $|p_1-p_2|$ to allow transfer. As a consequence, when $|p_1-p_2|<\delta_{min}$ or $|p_1-p_2|>\delta_{max}$ there is no transfer (i.e. $f(|p_1-p_2|) = 0$) and otherwise a constant fraction is transferred (i.e. $f(|p_1-p_2|) = \sigma$). Therefore, the parameter $0<\sigma<1$ is called the transfer efficiency, and $0\leq\delta_{min}<\delta_{max}$ are called the transfer thresholds.

## 3. RESULTS

### *3.1. MCF-7 proliferation and resistance.*

In 2005, Levchenko *et al.* have shown that several drug sensitive cell lines, including the well-established human breast adenocarcinoma cells MCF-7, are able to acquire multidrug resistance extragenetically, *via* a direct intercellular transfer of the resistance protein P-gp from a derived *mdr1*-transfected cell line. In the present work, the consequences of P-gp transfers in MCF-7 were studied by using, as a P-gp donor, a cell line variant obtained by selection for resistance to doxorubicine (DOXO) and therefore called MCF-7/DOXO. MCF-7/DOXO have previously been shown to overexpress P-gp, to resist to doxorubicin and to efflux P-gp-substrate fluorescent probes in a verapamil and cyclosporine A-sensitive manner [28].

Resistance phenotypes were first assessed both in MCF-7 and MCF-7/DOXO by determining growth rates in the presence of increasing DOXO concentrations (0, 0.1, 0.3, 1, 3 and 10 µM). As seen in Figure 1, without treatment, MCF-7 and MCF-7/DOXO displayed very similar growth rates, 0.6195 and 0.6328 per day, respectively. These values correspond to a doubling time of 26.8 and 26.3 hours, for sensitive and resistant cells, respectively. In the presence of DOXO (0.1-3 µM), the proliferation rates of MCF-7/DOXO were not significantly modified. By contrast, MCF-7 cells were strongly affected by the treatment, the growth rates passing through 0.0316 per day at 0.1 µM DOXO and decreasing very rapidly to -0.7492 per day with 3 µM DOXO.

### *3.2. Transfers of P-gp and efflux activity in co-cultures.*

In order to investigate potential transfers of P-gp between cells, 50 % of non P-gp-expressing parental MCF-7 cells were co-cultured with 50 % of their multidrug resistant counterpart MCF-7/DOXO. Obvious morphological differences allow distinction of MCF-7 from MCF-7/DOXO variants in the monolayer under phase contrast microscopy (Figure 2). After attachment to the tissue culture treated dish, MCF-7 were characterized by markedly birefringent membrane margins, while MCF-7/DOXO were more spread with dark edges and flat cell bodies. Following several days of growing in co-culture, we noticed a singular and stable spatial organization consisting in the formation of MCF-7 islets encased in a layer of MCF-7/DOXO.

A first approach used to detect a potential intercellular P-gp transfer was direct immunodetection of P-gp in living cells in the co-cultures, by using a phycoerythrin-conjugated monoclonal antibody directed against an epitope localized in an extracellular loop of the protein [29]. Membrane P-gp content was followed over time, from day 0 to day 6, by flow cytometry. As shown in Figure 3A, the peak corresponding to non P-gp-expressing MCF-7 at day 0 progressively drifted to the right, towards the regions of high P-gp levels, with time in co-culture. In addition, the total mass of P-gp remained unchanged over time for 10,000 analyzed cells (Figure 3C). Thus, it is suggested that that the distribution of P-gp within the cell population was reorganized with time but that the balance between P-gp biosynthesis and degradation remained constant, in average, in the samples.

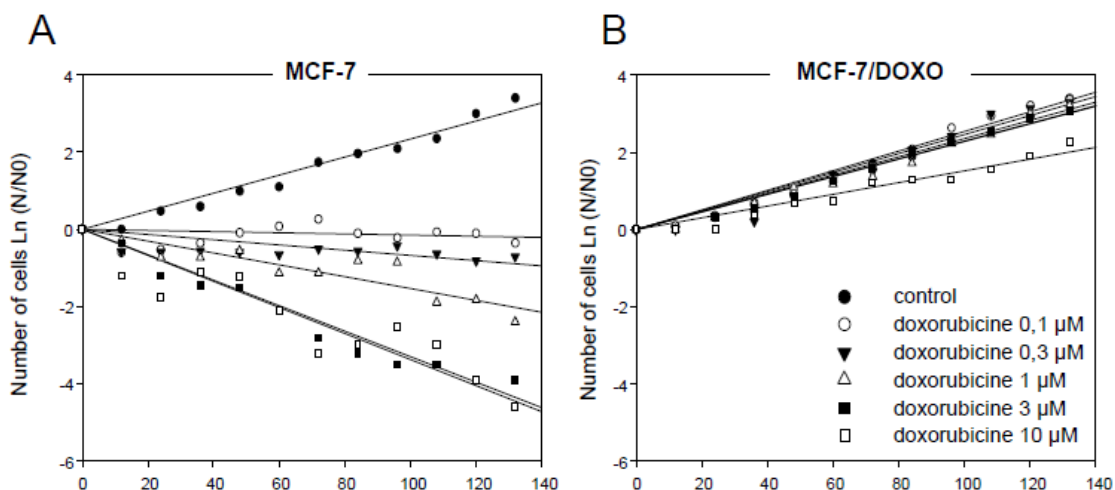

**Figure 1. Proliferation and resistance of MCF-7 variants in culture**.
Growth curves of MCF-7 (A) and MCF-7/DOXO (B) were established over 5,5 days. Cells were grown in the absence or presence of doxorubicin (0.1 to 10 µM, corresponding symbols given in the legend in B) and counted every 12 hours in a Malassez chamber. Cell counts are expressed as the logarithm of the cell numbers at various times (N) divided by the cell number at day 0 (N0).

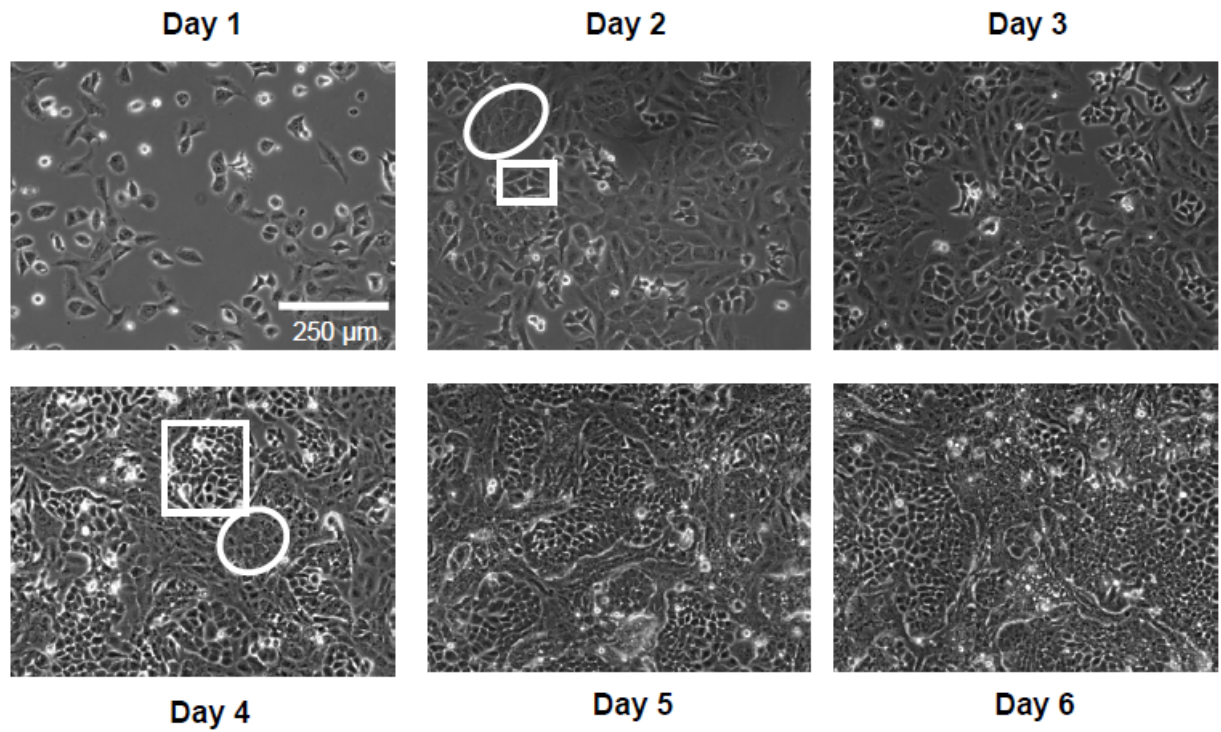

**Figure 2. Spatial organization of MCF-7 and MCF-7/DOXO in co-cultures.**
To obtain phase contrast micrographs of growing MCF-7 variants in co-cultures, dishes were seeded with a 50:50 mixture of MCF-7:MCF-7/DOXO at day 0. Morphological differences permit an immediate identification of each cell subpopulation. MCF-7 appeared birefringent and round (boxes) whereas MCF-7/DOXO are more flat and spread (ellipses). Note that the cells remained organized in well-delimited islets.

To investigate whether P-gp transfers generate a functional drug resistance, the ability of co-cultured cells to efflux calceinAM, a P-gp substrate probe which is a precursor of the fluorescent dye calcein was assayed, again by flow cytometry. In these experiments, cell fluorescence accumulation is therefore an inverse function of P-gp activity. Symmetrically to what observed with membrane P-gp content, the peak of fluorescence corresponding to initially sensitive cells, with a high calcein concentration, was also progressively shifted, but to the left, towards regions of less fluorescence accumulation (Figure 3B). These cells therefore acquired an aptitude to efflux the fluorescent dye. However, by contrast with Figure 3A, the activity distribution is additionally characterized by the appearance of a third subpopulation displaying an intermediate efflux activity. This middle peak increased in amplitude with time. Concurrently, the peak amplitudes corresponding to the subpopulations of MCF-7 with higher and, especially, lower calcein concentrations slightly declined. Interestingly, the total mass of activity (Figure 3D) also remained constant with time.

Although the progressive shift towards regions of higher membrane P-gp content, concurrently with the appearance of higher efflux activities, are observations that can be accounted for by effective P-gp transfer in co-cultures, the exact origin of these protein and activity redistributions had to be clarified. It should be first mentioned that, in our experiments, such phenomena are never observed in cells cultured separately. Nevertheless, in co-cultures, various cell interactions, secretion of soluble factors, could be responsible for microenvironmental regulations resulting in a decrease in P-gp expression in MCF-7/DOXO. Such phenomena could result in P-gp and activity redistributions, consistent with results observed in the preceding experiments. To investigate these points, it was necessary to distinguish and to follow both MCF-7 variants in co-cultures. Thus, parental sensitive MCF-7 cells were tagged with the persistent dye Cell Tracker Blue (ctbMCF-7) before mixture with MCF-7/DOXO and co-culture. As seen in Figure 4, upon one day of co-culture, a population of ctbMCF-7 appeared also positive to P-gp detection with the phycoerythrin-conjugated UIC2 mAb. This population appeared to be relatively stable over 3 days of co-culture, even though dilution of Celltracker blue within successive ctbMCF-7 generations weakened the signal and did not allow us to distinguish cell subpopulations further. In these experiments we also noticed continuous variation of phycoerythrin fluorescence in co-cultured ctbMCF-7, suggesting a progressive acquisition of P-gp by sensitive parental MCF-7, rather than a decrease of P-gp expression in MCF-7/DOXO. The possibility that dual positive events were, in fact, artifactual doublets of ctbMCF-7+MCF-analysis was triggered on the electronic Coulter-type cell volume (EV). Cell diameters in quadrants Q1, Q2 and Q4 matched to values corresponding to lone MCF-7 or MCF-7/DOXO cells (Figure 5).

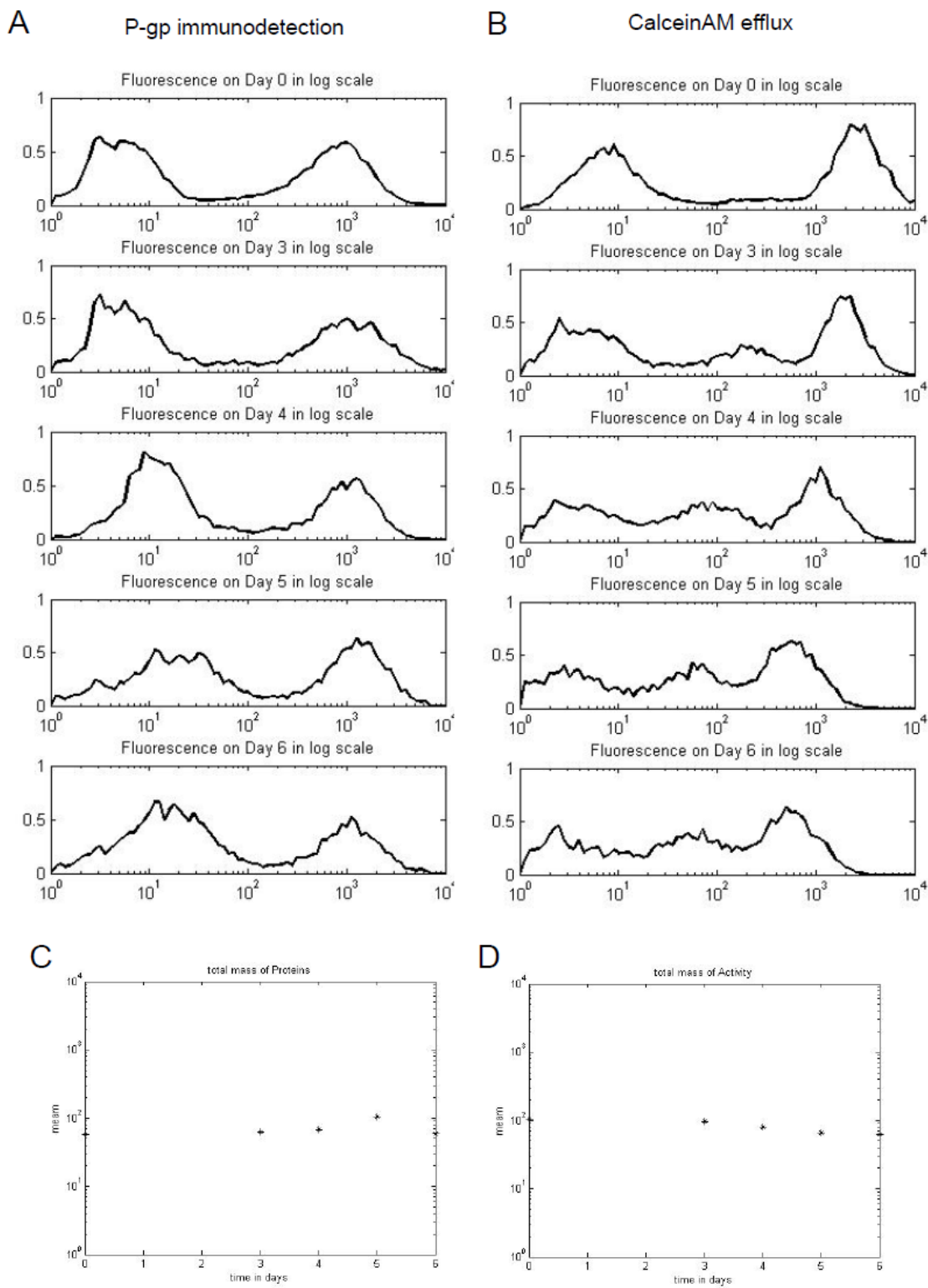

**Figure 3. Flow cytometry analysis of P-gp and efflux activity transfers between MCF-7 and MCF-7/DOXO variants of the human breast cancer cell line.**
A-B, A 50:50 MCF 7:MCF 7/DOXO cell mixture was seeded on cultures dishes at day 0 and co-cultured. P-gp expression was immunodected by using a PE-conjugated UIC2 monoclonal antibody (A) and P-gp activity was followed with calceinAM as a fluorescent probe (B) after 0, 3, 4, 5 and 6 days of co-culture. In both cases (A & B) a sample of 10 000 cells was analyzed. In order to reduce the stochastic fluctuations, normalized binned logarithmic histograms were built from all-events list-mode raw data (see methods). In A, the peak (on the left side) of MCF-7 expressing low levels of P-gp corresponds to the sensitive cells. The peak on the right side corresponds to MCF-7/DOXO resistant cells. In B, the peak (on the right side) corresponds to the low efflux activity of sensitive MCF-7 cells, and the peak on the right corresponds to the MCF-7/DOXO resistant cells. C-D, total mass (sum of collected fluorescence light of a sample) of P-gp (C) and efflux activity (D) in co-cultures have been computed through days 0, 3, 4, 5 and 6 for 10 000 events analyzed by flow cytometry.

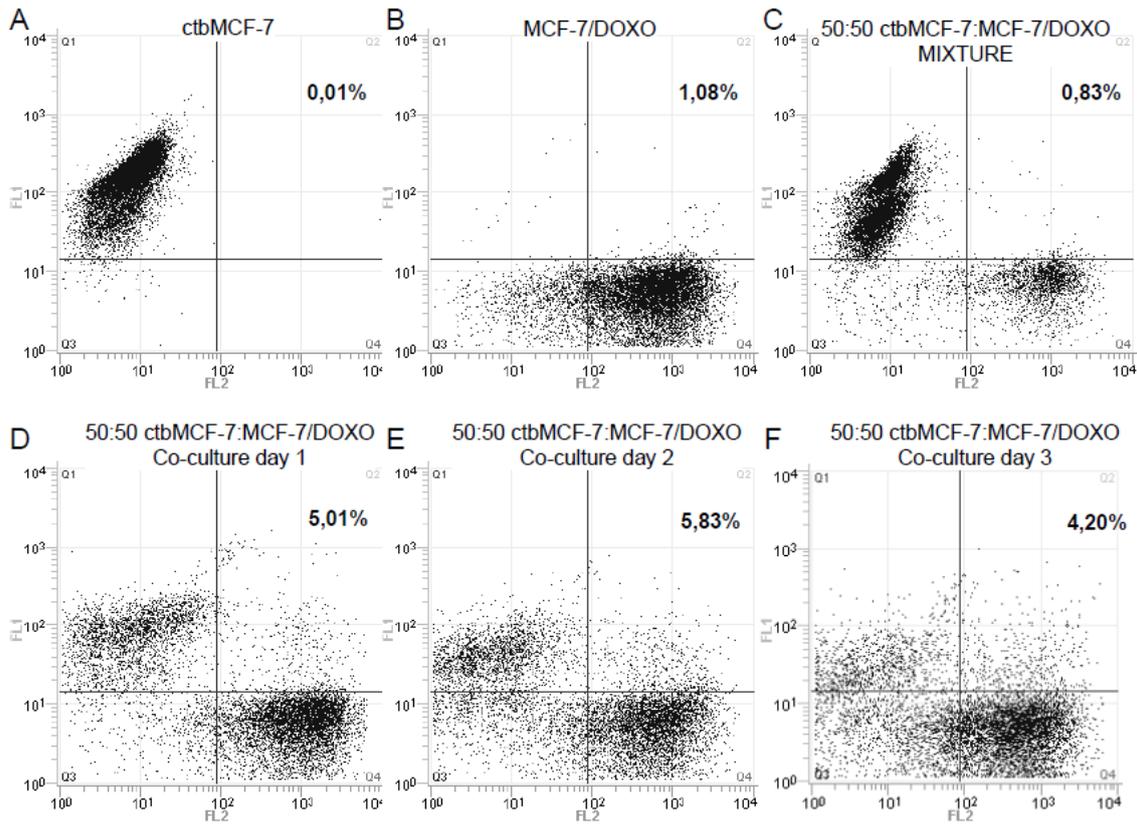

**Figure 4. Flow cytometry detection of P-gp transfers in tagged parental MCF-7.**
Parental sensitive MCF-7 were tagged with the persistent fluorescent probe CellTracker Blue (ctbMCF-7), prior to co-cultures. ctbMCF-7 alone, extemporaneous mixtures of 50:50 ctbMCF-7:MCF-7/DOXO or co-cultures at various times were analyzed after labelling with the PE-conjugated UIC2 monoclonal antibody. Scatter plots were obtained by quantifying CellTacker Blue content the FL1 channel and P-gp immunodetection in the FL2 channel. A, pure ctbMCF-7. B, pure MCF-7/DOXO. C, extemporaneous mixture of ctbMCF-7:MCF-7/DOXO. D-F, co-cultures obtained from 50:50 ctbMCF-7:MCF-7/DOXO, after 1 to 3 days. All analysis were performed with unchanged excitation light power and photomultipliers voltage settings. The quadrant limits were set in order to obtain less than 1 % of double positive cells (Q2) in the analysis of the sample C. Despite a progressive loss of FL1 signal, due to Celltracker blue dilution within daughter ctbMCF-7, a population of dually labelled cells appeared in the upper right quadrant. Percentages indicate the fraction of cells having a double positive labelling in quadrant Q2.

Taken together, these results strongly suggest the occurrence of a mechanism of transfer *i)* involving only a small fraction of the membrane content in protein of the donor cells but conferring a significant efflux activity to the recipient cells and *ii)* exhibiting a continuous exchange mode simply dependent on the gradient, *i.e.* the net difference in P-gp expression levels separating the cells under concern. In order to test these hypotheses and to evaluate their consequences in the dynamics of multidrug resistance within a cell culture, a mathematical transfer model was developed.

***3.3. Estimation of transfer parameters.***

To quantify the transfer of P-gp activity, which is assumed to result in a gain of resistance, we have used a mathematical model, presented in the material and methods section and derived from Hinow *et al.* (2009), with no-random noise (*i.e.* $\varepsilon=0$), and without efflux activity drift (*i.e.* h=0). These terms were neglected, since a short-term (6 days) period was considered here, while these terms were introduced to describe a long-term evolution of the cell culture (*i.e.* a period of months). In addition, in the absence of cytotoxics and according to Figure 1, the growth rate is independent of the P-gp content. Herein, the growth rate was thus assumed to be identical for MCF-7 and MCF-7/DOXO in co-culture ($\rho=0.63$ day$^{-1}$).

The transfer rate τ, the transfer efficiency f, and the thresholds $\delta_{min} < \delta_{max}$ were estimated by using intensive parameter estimation computations. The conditioned flow cytometry activity distribution at day 0 was used as an input for the numerical simulation of the model. The best fit to data distribution at day 6 was estimated by using the least square minimization method over 100,000 different values of the parameters (Figure 6). Conclusively, the model is coherent with the existence of cell-to-cell resistance transfers, since the best fit of the parameters gives τ=0.4, and f =0.2, corresponding to non-null transfer rate and efficiency. The parameters of the model, the respective methods used to determine their numerical values and the associated symbols are given in Table 1.

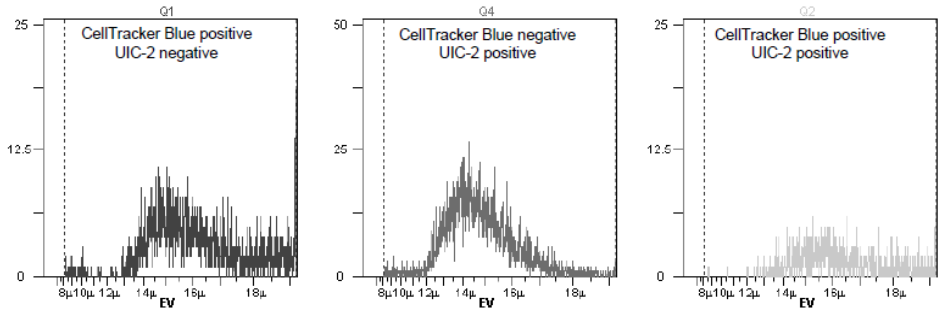

**Figure 5. Distribution of cell diameters in parental sensitive, multidrug resistant and transferred MCF-7 populations.**
The histograms correspond to cell diameters in quadrants Q1 (CellTracker Blue positive cells, left), Q4 (UIC2 positive cells, middle) and Q2 (double positive cells, right) of Figure 4. Quantitative values give cells volumes of 1984 ± 51 µm3 for parental sensitive MCF-7, 1776 ± 41 µm3 for multidrug resistant MCF-7/DOXO and 2398 ± 39 µm3 for transferred MCF-7 (mean ± coefficient of variation).

Fitted parameters seem to indicate that some limitations, or constraints, occurred for transfers between cells in our experimental conditions. First, the transfers were not homogeneous over the whole assortment of activities expressed by cells within the co-cultures. In this respect, a threshold $\delta_{min}$ distinct from 0 suggests that transfers did not occur or were not detected when P-gp activities of the cells were to close. In addition, since the threshold $\delta_{max}$ is not equal to the maximal value $10^4$, transfers did not arise either in case of extreme activity differences. This implies the occurrence of a permissible range of activities governing the optimal transfers between donor and recipient cells. Moreover, a transfer rate of $\tau=0.4$ means that transfer events occurred in average every $1/\tau= 2.5$ days for each cell in our experiments. Finally, the quantity of transferred activity globally corresponds to a fraction $f=0.2$ of the difference between the activity of the donor and recipient cell.

sensitive and resistant cells, simulations were carried out by using our mathematical model of activity transfers, fed with biological data.

Herein, the simulations were built with basic parameters determined *in vitro*, in particular the growth rates of monolayers and the transfer parameters. Therefore, cell proliferation was artificially high in abstraction of any physiological environment and the growth curves obtained by simulation should not be considered as data clinically exploitable. Likewise, pharmacokinetics of doxorubicin (such as elimination half-life, drug clearance, volumes of distribution) was ignored. The simulated treatment was simplified into a bolus of 1 µM doxorubicin that reaches the maximum concentration in the extracellular compartment with no delay and maintains the same level over 10 days. Then, the treatment was interrupted for 4 days (*i.e.* doxorubicin concentration is zero). The treatment cycled every 14 days for 4 cycles.

As shown in Figure 7A, the number of pure sensitive MCF-7 in a simulated culture declined progressively in the presence of DOXO during the first 10 days of a treatment cycle and, thereafter, increased

| Symbol | Interpretation | Value | Units | Method |
|---|---|---|---|---|
| h | P-gp activity Drift | 0 | day$^{-1}$ | set |
| ρ | Growth rate of cells in absence of drug | 0.63 | day$^{-1}$ | measured |
| c | Drug concentration | 0 or 1 | µM | set |
| ε | Stochastic drift of P-gp activity | 0 | - | set |
| τ | Rate of transfer of P-gp activity | 0.4 | day$^{-1}$ | fitted |
| f | Activity transfer efficiency | 0.2 | - | fitted |
| $\delta_{min}$ | Minimum threshold for transfers | $10^{1.4}$ | fluorescence unit | fitted |
| $\delta_{max}$ | Maximum threshold for transfers | $10^{1.9}$ | fluorescence unit | fitted |

**Table 1. List of the model parameters, their significations, values and symbols.**
The column method refers to the approach employed to determine their respective numerical value

### 3.4. Consequences of resistance activity transfers on model response to chemotherapeutic cycles in co-cultures.

Fast acquisition of P-gp activity by doxorubicin-sensitive MCF-7 in a population of growing tumoral cells may be of first relevance to determine the dynamics of response to chemotherapy. Thus, to test the significance of P-gp transfers within a co-culture of during the 4-days interruption at rates $\rho(p_{min})$ corresponding to the values previously determined (Figure 1). To follow the respective populations of sensitive and resistant cells within the simulated co-cultures, a particular attention should be paid to the criterion used to define what exactly a sensitive cell is (resistant cells being considered as the complement). Herein, we defined sensitive cells as MCF-7 having a negative growth rate in the presence of 1 µM doxorubicin. The growth rate ρ is supposed to vary only as a function of the overall P-gp activity $p$ (intrinsic activity and activity acquired or lost by transfers), as presented in supporting Figure 7B. As a consequence, transferred sensitive cells with enough P-gp activity to have a positive growth rate will be considered as resistant cells.

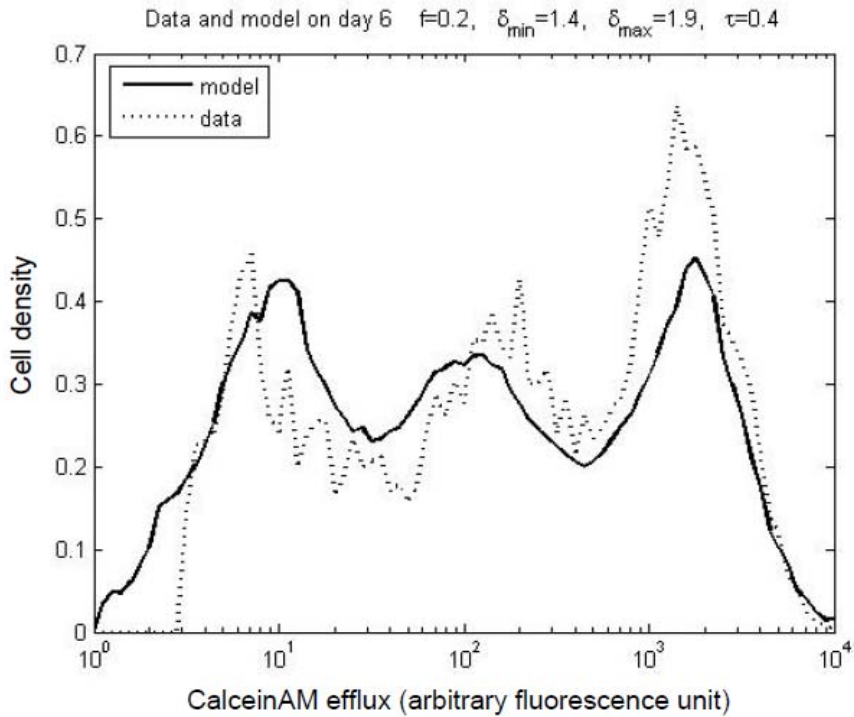

**Figure 6. Estimation of parameters for the transfer model.**
This graph corresponds to the distribution of P-gp activity at day 6 of co-culture. The initial distribution of P-gp activity at day 0 (not shown here) was obtained from a mixture of 50:50 MCF-7:MCF-7/DOXO cells analyzed by cytometry as in Figure 4. The transfer model was run over 6 days. The dotted curve corresponds to the efflux activity of the co-culture measured at day 6, and the solid curve corresponds to the distribution of activity derived from model. The fitting parameters are given above the curves and were obtained by least squares minimization.

The evolution of the simulated co-cultures was extremely dependent on the initial distribution of cells, *i.e.* the relative proportions of cells of with different P-gp contents. Again, as for the estimation of transfer parameters, the biological data distribution in term of efflux activity $p$ at day 0, was used as an input template for computer simulations. However, the simulations were started with inocula of $10^4$ cells (corresponding to an approximate mass of 0,01 mg) comprising only a small proportion of resistant cells, 0.1 % (*i.e.* 10 cells, Figures 8A&B), or 1 % (*i.e.* 100 cells, Figures 8C&D). For each condition, the numbers of sensitive, resistant and total cells were plotted as a function of time in two situations: without or with cell-to-cell P-gp activity transfers.

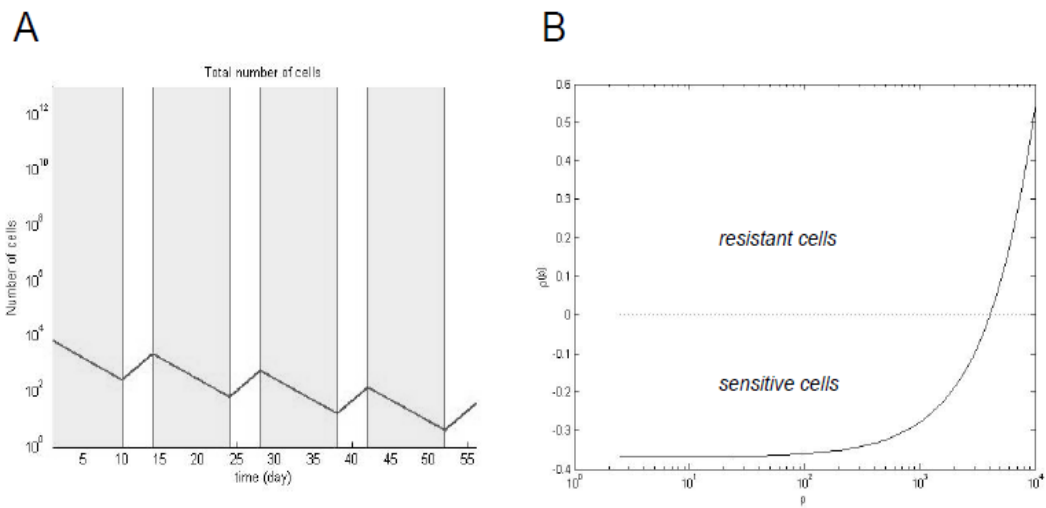

**Figure 7. Growth rate of sensitive cell as a function of P-gp content.**
A, computer simulation giving the number of sensitive MCF-7 versus time during a chemotherapy regimen, cycling every 14 days, with 10 days of 1 μM doxorubicin treatment followed by 4 days of interruption, for 4 cycles. The initial simulated tumor consisted in 104 MCF-7 having growth rates corresponding to ρ(pmin). B, plot giving the growth rates ρ as a function of the P-gp-driven calceinAM efflux activity p for cells exposed to 1 μM doxorubicin. The function describing ρ(p) is used in the simulations presented Figure 8. All cells having a negative growth rate are considered as sensitive MCF-7.

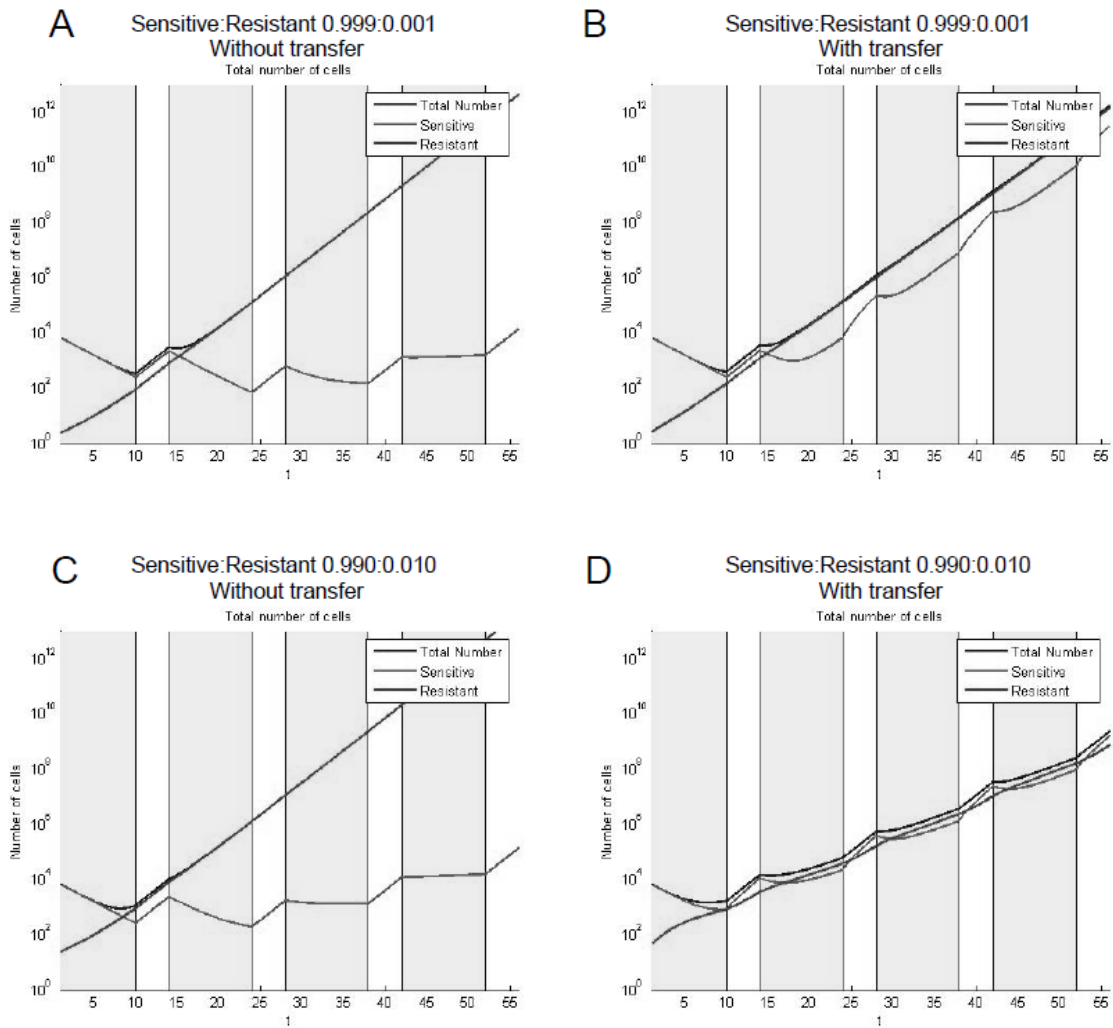

**Figure 8. Computer simulation giving the consequences of cell-to-cell P-gp activity transfers in co-cultures exposed to doxorubicin, according to the mathematical model.**
A chemotherapy regimen is simulated by cyclic exposure of an inoculum of 104 cells to 1 µM doxorubicin. The treatment cycled every 14 days, with 10 days of doxorubicin treatment followed by 4 days of interruption, for 4 cycles. The initial proportions of resistant MCF-7/DOXO vary form 0,1 % (A & B) to 1 % (C & B). Simulations were performed in situations where transfers were abolished (A & C) or permitted (B & D) with the parameters determined in Figure 6. For each condition, the numbers of sensitive, resistant and total cells are plotted versus time. Sensitive MCF-7 are defined as cells having negative growth rate in the presence of 1 µM doxorubicin (see Figure 7).

In the absence of transfer (Figure 8A&C), resistant cells grew linearly, and represent after few days the majority of the cells in the co-cultures. The number of sensitive cells decreased during the treatment and increased during the interruption phases. In these simulations, the initial distribution of sensitive cells was based on low efflux activities analyzed by flow cytometry. As a consequence, the population of sensitive cells did not correspond to a single ray but was rather represented by a bell curve shape encompassing a spectrum of efflux activities, from zero to several arbitrary fluorescent units. Consequently, from a cycle to another, the sensitive cells having the less negative growth rates were selected by the treatment. For that reason, the population of sensitive cells decreased to a minimum and, then, increased again by growing during treatment interruptions while the growth rates converged to zero during the doxorubicin phases.

In the presence of transfers (Figure 8B&D), results differed from above and were greatly influenced by the initial number of resistant cells in the inoculum. In fact, with transfers during treatment, growth rates of resistant and sensitive cells converged to an intermediate value, due to a loss of activity by resistant combined with a gain of activity by sensitive cells. With initial conditions corresponding to 0.1 % or 1 % of resistant cells, after an initial phase during the first treatment, the growth rates of the cell population tend to become analogous. The main observation is that transfers significantly decreased the growth rate of the whole cell population exposed to cytotoxics, especially when the co-culture is initiated with a relative large proportion (1 %) of resistant cells.

## 4. DISCUSSION

In the present work, we observe effective transfers of P-gp and efflux activity from a P-gp overexpressing MCF-7 variant, selected for its resistance towards doxorubicin, to the parental sensitive cell line. Our results confirm the work of Levchenko *et al.* (2005) conducted *in vivo* and also *in vitro* on several cell lines, including *MDR1*-transfected MCF-7. In 2008, the ability of stromal ovarian cells to confer chemoresistance, through heterocellular interactions and membrane exchanges, has also been reported and identified as "oncologic trogocytosis" in reference to analogous cell-to-cell protein transfers occurring in "immunological synapses" involving T cells [24]. In 2009, a functional transfer of P-gp between acute MDR+ leukemic cells cultured in suspension and their co-cultured sensitive counterpart variant has also been found [25]. To our best knowledge, the present work and the three above cited references are the first reports describing extragenetic direct cell-to-cell transfers of P-gp. Among these lines, it should be mentioned that an intercellular transfer of the drug resistance phenotype was first reported indirectly in a pioneering study of Frankfurt *et al.* [30].

The present data highly reinforce the fact that the P-glycoprotein, a 170 kDa polypeptide consisting of 1280 amino acids that spans 12 times the cell membrane [31], can be exchanged from a donor tumoral cell to a recipient one, keeping its ability to efflux drugs and, thus, conferring the MDR phenotype. In addition to the discovery of the ability, for cells, to transfer huge integral membrane proteins, these important findings also reveal the occurrence of an additional extragenetic mechanism to acquired chemoresistance, in the absence of pressure selection. The present data also increase the list of cells capable of protein transfers.

The particular spatial organization of MCF-7 in co-cultures has certainly severe consequences on the contact surfaces between sensitive and resistant cells. As seen under microscopic observation, cells were not organized as a homogeneous mixture of well-separated cell variants, but rather in delimited islets of sensitive settling aside from resistant cells. As a result, transfers between the most separated populations in term of chemoresistance (full sensitive *versus* full resistant cells) may occur only at the frontier of islets, keeping in mind that transfers can also arise between cells of intermediate P-gp levels. From this, it is assumed that the transfer process is a rather complicated phenomenon in which spatial structures bring limitations. As a matter of fact, our approach to quantify the parameters of transfers by using biomathematical modelling, points out some constraints, revealed by the parameters fitted to the biological data. The mathematical model has been constructed to *i)* be applicable to a population of interacting cells in proliferation, *ii)* to consider the cell population as a continuum density structured by the quantity $p$ of transferable P-gp activity and *iii)* to have rules governing transfer. The rules of the transfer process have been delineated and discussed in detail in a companion paper [26]. The better fit was obtained by using the activity transferred from resistant to sensitive cells, undirected and directed shifts being neglected. Herein, we consider that P-gp activity is a better estimation of therapy resistance as compared to P-gp expression, because activity is directly responsible for resistance to cytotoxic drugs. Moreover, this choice allows us to use directly the model to simulate the response of a co-culture to chemotherapeutic cycles (see below). Nevertheless, a correlation between P-gp levels and activity has been described in several cell lines [15, 32, 33]. Parameters fitting revealed quantitative aspects of P-gp activity transfer, in particular a transfer rate $\tau$ corresponding to one event every 2.5 days, a value to be compared to the doubling time of MCF-7 close to 1 day. In addition, the calculation of the average activity and the average level of P-gp expression over the time in a co-culture showed that theses values remain stable throughout the experiments. We conclude that there is neither significant gain nor loss of efflux activity and P-gp expression during the co-culture experiments. As a consequence, the changes in the fluorescence distributions (for both P-gp and activity) can be attributed to transfers, which are not rare phenomena compared to the cell cycle duration. However, in a culture of growing cells transferring P-gp, the topology of the exchange network is constantly changing. This aspect is not taken into consideration in the transfer model which does not adapt rules of transfer as a function of time. The estimation of the transfer rate $\tau$ is therefore a coarse estimation that should be re-evaluated in future works including spatial dynamics.

The model gives also a transfer efficiency $f = 20$ %, corresponding to the amount of activity transferred each time with regard to the absolute activity difference between the donor and recipient cells, and two transfer thresholds, that can be viewed as bottom and top activities delimiting authorized transfers. These parameters can be related to the life span of potential transfer vectors, that are constructed and destroyed for instance, or to the size of single P-gp cargos. Concerning this point, it has been shown that high density P-gp microdomains alter local properties of lipid membrane environment and promote P-gp clustering [34]. In this respect, biological data as well as modelized transfer parameters suggest an organized quantal transfer of P-gp. Several intercellular carriers have been proposed to be involved in P-gp transfers. In adherent cells lines, close contacts between donor and recipient cells have been shown to be required [23, 24]. On the other, in a model of liquid tumor, P-gp containing microparticles released by resistant cells and bound to sensitive cells, have been isolated [25]. The possibility that both mechanisms actually co-exist in our cell line is examined in an independent study [27].

The model was used to simulate and to quantify the consequences of P-gp activity transfers on the cell distribution within a growing co-culture during exposure cycles to cytotoxics. More precisely, the model was fed with biological data corresponding to sensitivity of MCF-7 variants to doxorubicin. Doxorubicin is among the preferred agents used in chemotherapy regimens for preoperative breast tumor reduction, adjuvant treatment after a first line surgical lumpectomy/mastectomy and radiotherapy of invasive breast carcinoma or systemic treatment of recurrent or metastatic breast cancers [35]. Doxorubicin is administrated alone or in combinations with other cytotoxics. In the present simulations, the concentration of doxorubicin was set at 1 µM, a value corresponding to the weekly supplementation of MCF-7/DOXO culture medium. This concentration is also close to the levels of doxorubicin found in tumors under chemotherapy, reported to vary in the 0,1-10 µM range depending on the time separating the injection to the measures [36-38].

The major consequence of P-gp activity transfers is that subpopulations of sensitive and resistant cells were no longer compartmentalized. *Via* P-gp activity transfers, flows of cells depopulate and repopulate resistant and sensitive cells. This was obvious once the second cycle of treatment began, where the population of sensitive cells increased by adjunction of cells coming from resistant cells, which lose P-gp activity. Conversely, sensitive cells acquired P-gp activity and, thus, developed progressively positive growth rates during the treatment phases. When the number of donor cells was increased up to 1 %, the resulting exchanges of P-gp caused a significant loss of activity for the resistant cells. Acquisition of P-gp activity was therefore slowed down for sensitive cells. As a result, in that case, the numbers of sensitive and resistant cells within the co-culture converged (became close) with time and the overall growth rate of the cell population was reduced (compared to all the other situations), giving a co-culture with fewer cells at the end of the four chemotherapeutic cycles. In our model, the gain of P-gp is central for survival and growth of non P-gp expressing cells whereas the loss of P-gp by resistant cells has predominant effects on the overall growth rate of the simulated tumour. We also conclude that the initial proportion of resistant cells in a population has a great influence on the growth of the whole population in the presence of transfer.

The present study sheds new light on fundamental features underlying extragenetic acquisition of multidrug resistance *in vitro*, with a particular attention given to quantitative aspects. Globally, P-gp transfers led to an integration of the responses to the drug across the cell population and a 'collective' phenotype different than the sum of parts. However, MCF-7/DOXO represents an extreme case of P-gp overexpression and fast growing cells. Future work should be conducted with cells expressing intermediate amounts of efflux activity, since transfers have been demonstrated to be dependent on the P-gp levels [23]. Moreover, an improved mathematical model taking into account evolving 2D or 3D spatial considerations has to be developed together. Physiopathological implications of cell-to-cell protein transfers may be crucial, especially if other membrane proteins and various cell types are concerned.

## 5. ACKNOWLEDGMENTS

This work was supported by grants of the federation FED 4116 SCALE (SCiences Appliquées à L'Environnement). Jennifer Pasquier was a recipient for a fellowship from the Conseil Regional de Haute-Normandie The authors are indebted to Pr. Jean-Pierre Marie (Hôtel Dieu, Paris, France) for providing MCF-7/DOXO and PSC833.